\documentclass[a4paper,10pt]{article} 

\usepackage[english]{babel}
\usepackage{amsmath,amsthm,amssymb,amscd,amsfonts}
\usepackage{graphicx}
\usepackage[english]{babel}

\newtheorem{teo}{Theorem}[section]

\newtheorem{lemma}[teo]{Lemma}

\newcommand{\f}{\frac}

\title{Can a traveling wave connect two unstable states? \\
 The case of the nonlocal Fisher equation}
\author{Gr\'egoire Nadin \thanks{UPMC, CNRS, UMR 7598, Laboratoire Jacques-Louis Lions, F-75005, Paris. Email: nadin@ann.jussieu.fr}
\and Beno\^ \i t Perthame \footnotemark[1] \thanks{
INRIA Paris-Rocquencourt, Equipe BANG. Email: benoit.perthame@upmc.fr}
\and Min Tang  \footnotemark[2] \thanks{Email: mintang@ann.jussieu.fr}
}

\date{\today}

\begin{document}
\maketitle

\begin{abstract}
This note investigates the properties of the traveling waves solutions of the nonlocal Fisher equation. The existence of such solutions has been proved recently in \cite{BNPR} but their asymptotic behavior was still unclear. We use here a new numerical approximation of these traveling waves which shows that
some traveling waves connect the two homogeneous steady states $0$ and $1$, which is a striking fact since $0$ is dynamically unstable and $1$ is unstable in the sense of Turing.
\end{abstract}

\noindent {\bf Key-words}  Nonlocal Fisher equation, Turing instability, traveling waves.
\\
\noindent {\bf AMS Subject Classification}:  35B36, 92B05, 37M99.
\section{Introduction}
\label{sec:intro}

For the semilinear equation
\begin{equation}\label{eq:semilnear}
\partial_tu- \partial_{xx}u =f(u),  \qquad \quad f(0)=f(1)=0,
\end{equation} a traveling wave can connect the two steady states $u\equiv0$ and $u\equiv 1$ in various  situations. This is the case when $0$ is unstable and $1$ is stable  (Fisher/monostable). This is also the case when $0$ and $1$ are both stable (Allen-Cahn/bistable). It is known that these waves are  attractive and are obtained as the long time limit of the dynamics (\ref{eq:semilnear}) associated with compactly supported initial data (see \cite{KPP}). When $0$ and $1$ are unstable, in between there is a stable steady state of $f$ which prevents any traveling wave to exists.

For systems and non-local equations, the classification becomes more complicated because a steady state can be Turing unstable, which means that $1$ is unstable with respect to some periodic perturbations in a bounded range of periods. In this note we consider the nonlocal Fisher equation (\ref{eq:nonlocal}), the simplest to produce Turing instability.
\begin{equation}\label{eq:nonlocal}
\partial_tu=\partial_{xx}u+ \mu u(1-\phi\star u), \qquad  \qquad \phi\star u(x)=\int_{\mathbb{R}}u(x-y)\phi(y)dy,
\end{equation}
with
\begin{equation}\label{eq:phi}
\mu >0, \qquad  \phi(x)\geq 0,\quad\phi(0)>0,\quad\int_{\mathbb{R}}\phi(x)\,dx=1.
\end{equation}
The steady state $1$ can be Turing unstable when the Fourier transform of $\phi$ changes sign and  $\mu$ is large enough (see \cite{FKK, GVA, BNPR}). This creates several differences with the monostable or bistable equations.

First of all, one can observe that the solution of (\ref{eq:nonlocal}) associated with a compactly supported initial datum does not converge, for large times, toward the traveling wave but it converges towards a more complicated structure (see \cite{FKK}, \cite{GVA} and \cite{Gourley}).
The numerical simulation presented in b) of Figure \ref{fig:nonlocal1} shows that the solution of the evolution equation converges to a pulsating front, that is a function $u(x-\sigma t, x)$ which is periodic in its second variable. These types of fronts typically arise in the framework of reaction-diffusion equations with periodic coefficients (\cite{BH}, \cite{Shigesada}). In a) of Figure \ref{fig:nonlocal1}, the solution seems to converge to the superposition of a traveling wave and a pulsating front, with two different speeds. These periodic patterns are a symptom of Turing instability on the full line $\mathbb{R}$.
\begin{figure}
\begin{center}
a)
\includegraphics[width=0.45\textwidth]{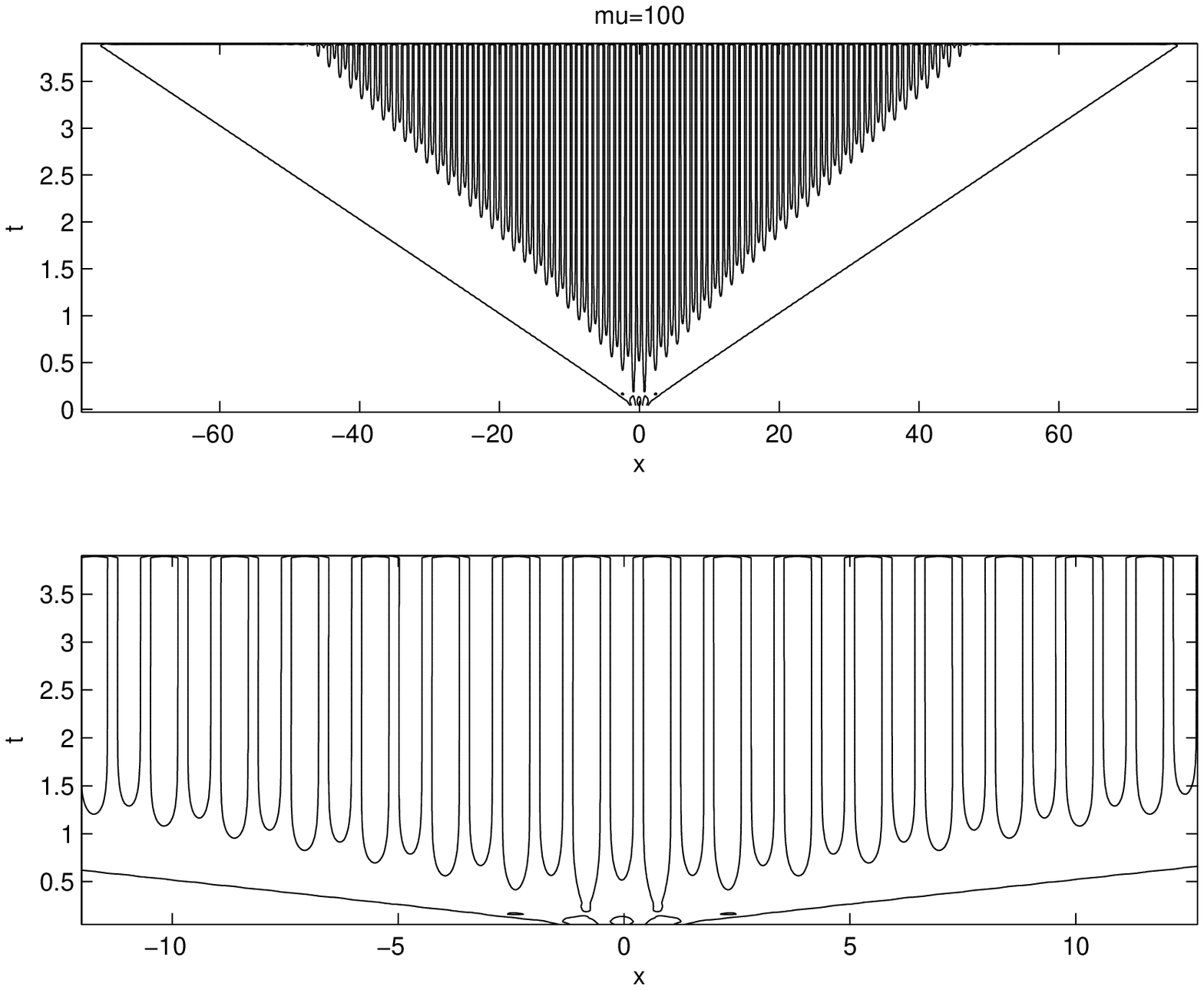}
\includegraphics[width=0.45\textwidth]{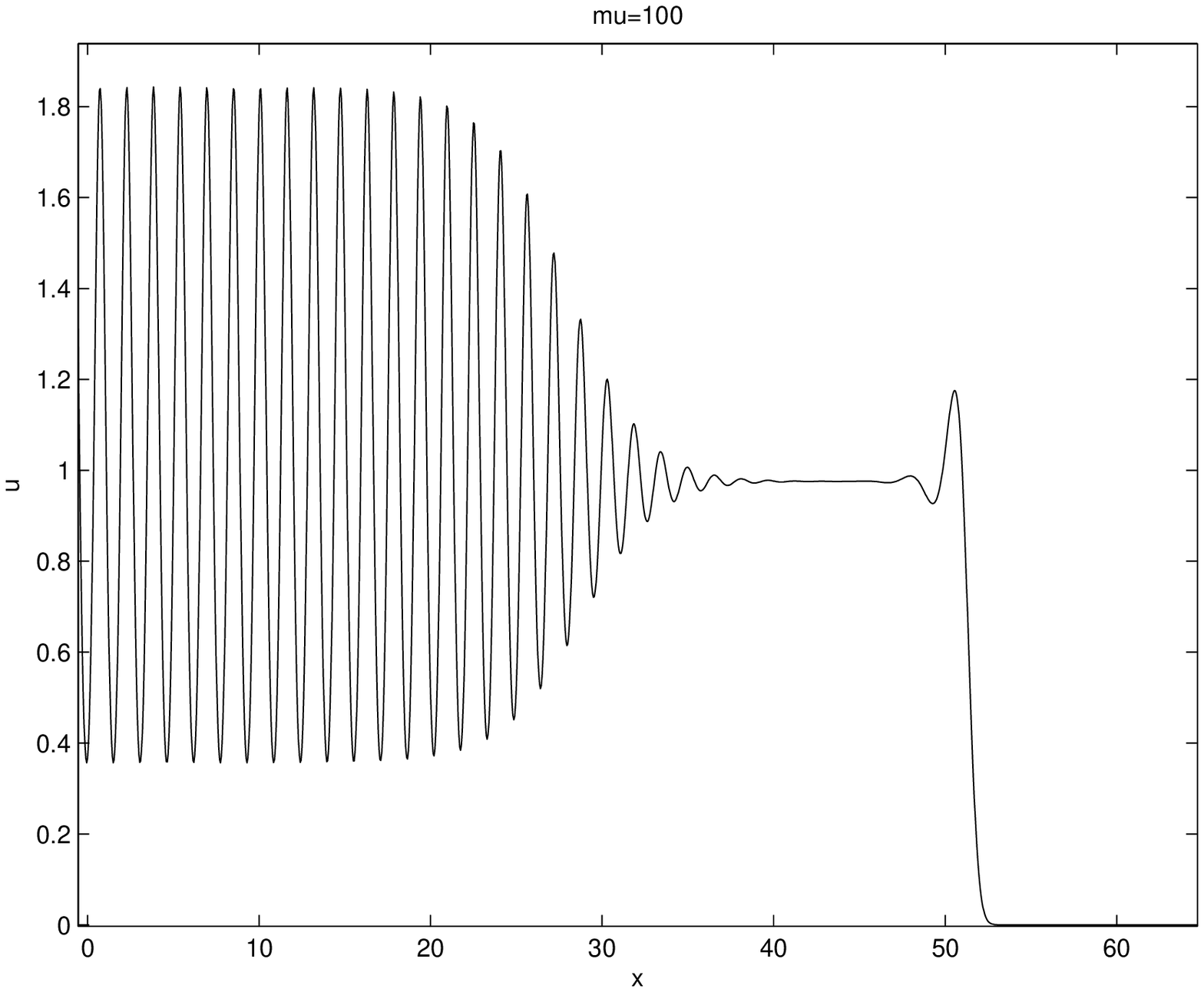}
b)\includegraphics[width=0.45\textwidth]{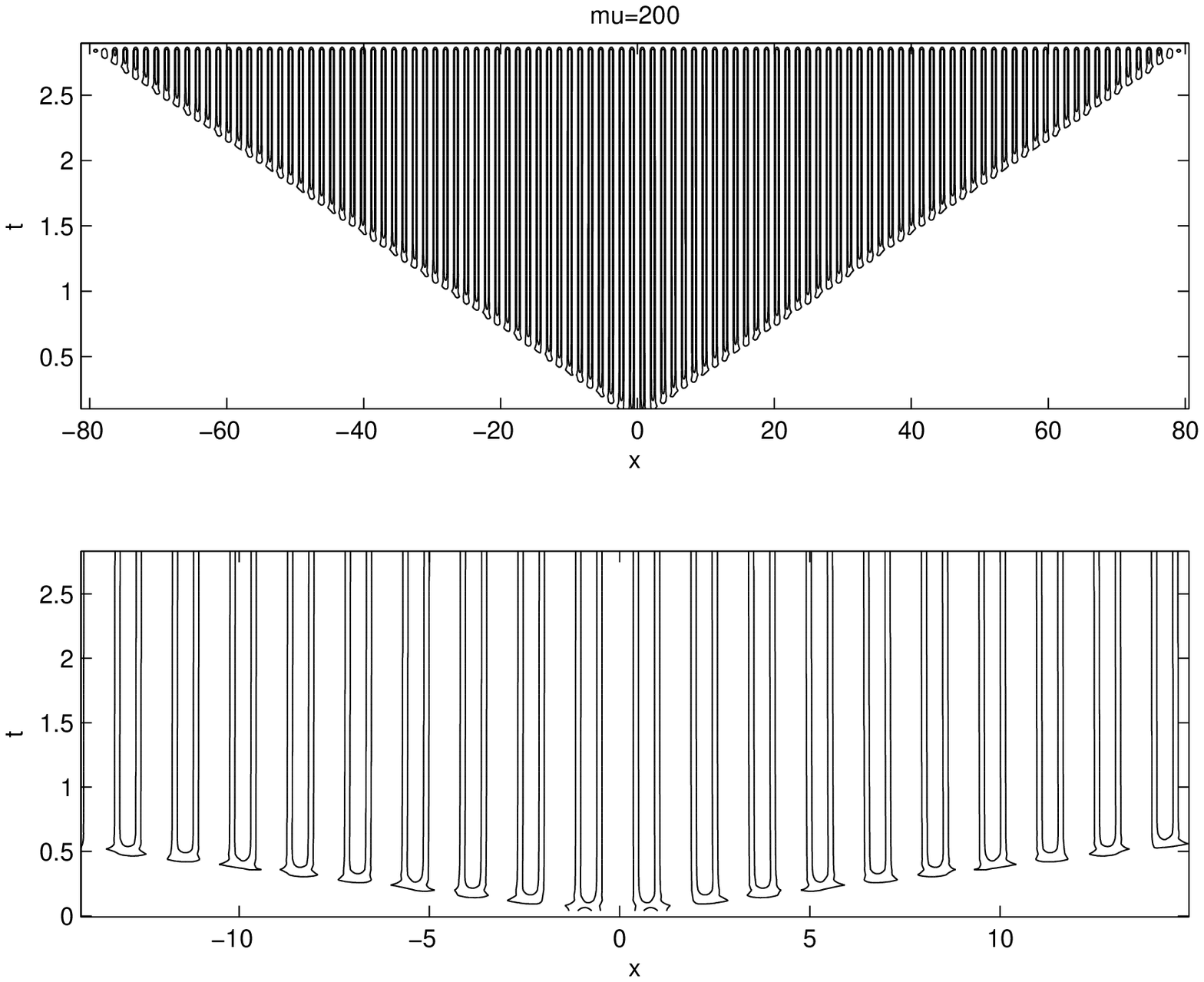}
\includegraphics[width=0.45\textwidth]{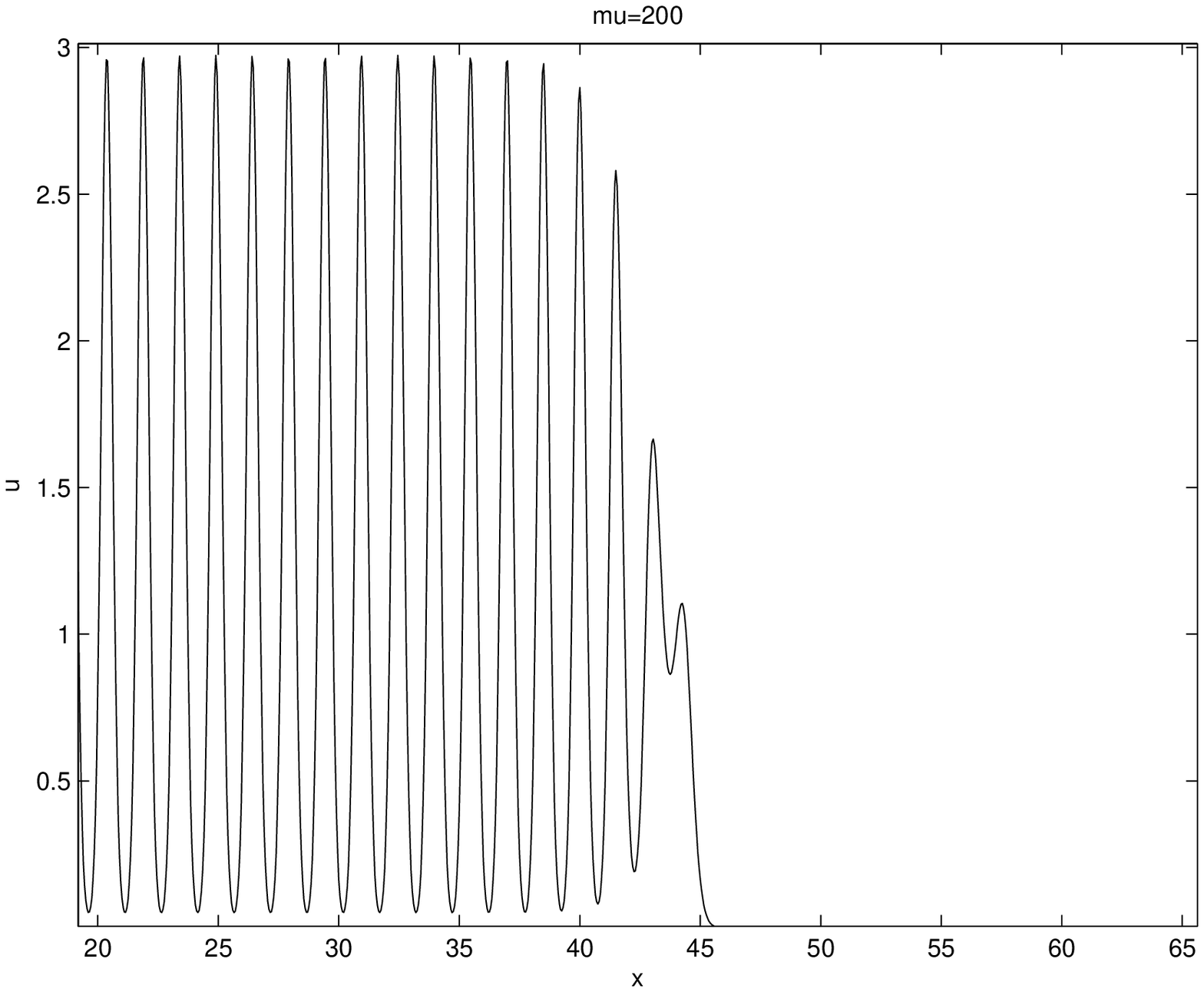}
\vspace{-2mm}
\caption{Numerical simulations of the time evolution  for the nonlocal Fisher/monostable equation (\ref{eq:nonlocal}) with kernel (\ref{eq:phi2}). The computational domain is $[-80,80]$. Left: the isovalues show that it is not a traveling wave but a more complicated structure; the bottom subplots are zooms of the top subplots. Right: the function $v^L$ connects the (dynamically) unstable state $0$ with a periodic tail at $x=-\infty$. Two values of $\mu$ are used a) $\mu=100$; b) $\mu=200$.}
\label{fig:nonlocal1}
\end{center}
\end{figure}

Secondly, it is proved in \cite{BNPR} that a traveling wave $u(x,t)=v(x-\sigma t)$ always exists for all $\sigma\geq 2\sqrt{\mu}$, with a generalized formulation
\begin{equation}\label{eq:nonlocalTW}
\left\{
\begin{array}{l}
\sigma\partial_x v+\partial_{xx}v+\mu v(1-\phi\star v)=0, \qquad x \in {\mathbb{R}}, \\
\liminf_{x\to-\infty}  v(x)>0,\qquad v(+\infty)=0.
\end{array}\right.
\end{equation}
The authors were only able to obtain, due to the nonlocal effect, the weak boundary condition at $x=-\infty$ rather than the expected condition $u(-\infty)=1$.
When $\mu$ is small enough or when the Fourier transform of the kernel ${\mathcal F} ( \phi) $ is positive everywhere, the traveling wave connects $0$ to  $1$. But this leaves open the question to know whether for $\mu$ large, the traveling wave solution of (\ref{eq:phi2}) connects the (dynamically) unstable state $0$ to a stable periodic state or to the Turing unstable state  $1$.
Also, when ${\mathcal F} ( \phi) $ can take negative values, it is proved in \cite{BNPR} that for $\mu$ large enough monotonic traveling waves cannot exist.

In the sequel, we will firstly focus on the case $\sigma=2\sqrt{\mu}$ and
\begin{equation}\label{eq:phi2}
\phi (x)=\frac {1}{2}\mathbf{1}_{[-1,1]}(x).
\end{equation}
Then we also test two other cases
\begin{subequations}\label{eq:phinew}
\begin{equation}
\phi (x)=\Big(-x^2+\beta^2\Big) \mathbf{1}_{[-\beta,\beta]},\qquad \beta=\big(\frac{3}{4}\big)^{\frac{1}{3}}
\label{eq:phinew1}\end{equation}
\begin{equation}\phi (x)=\big(\frac{1}{4}+\frac{1}{2}|x|\big)\mathbf{1}_{[-1,1]}(x)
\label{eq:phinew2} \end{equation}
\end{subequations}
In all these cases, ${\mathcal F} ( \phi) (\xi)$  takes negative values. For instance, for (\ref{eq:phi2}), ${\mathcal F} ( \phi) (\xi)=\sin \xi/ \xi$.
Finally, to complete the tests, the traveling waves for the Gaussian kernel
\begin{equation}\label{eq:phi1}
\phi(x)=\frac{1}{\sqrt{2\pi}}\exp\Big(-\f{x^2}{2}\Big),
\end{equation}
which has positive Fourier transform, are displayed.

In section \ref{sec:algorithm}, we develop a specific algorithm which allows us to build the traveling wave (\ref{eq:nonlocalTW}) and not the pulsating front. Then, in section in \ref{sec:results}, we perform numerical  simulations to decide which alternative holds true.

\section{The algorithm}
\label{sec:algorithm}

Numerically one can only solve the problem on a bounded domain of length $L=x_r-x_l$ (this is also the analytical construction in \cite{BNPR})
\begin{equation}\label{eq:nonlocalTWL}
\left\{
\begin{array}{l}
\sigma^L\partial_x v^L+\partial_{xx}v^L+\mu v^L(1-\phi\star v^L)=0,   \qquad  x_l<x <x_r,
\\
v^L(x_l)=1,\quad v^L(x_r)=0,\quad v^L(0)=\epsilon,
\end{array}\right.
\end{equation}
 The convolution is computed by extending $v^L$ by 1
on $(-\infty,x_l)$ and $0$ on $(x_r,+\infty)$. The parameter $\epsilon$, small enough, is needed for technical reasons but intuitively its value has to be below the oscillations observed in Figure \ref{fig:nonlocal2}.

Our algorithm for solving (\ref{eq:nonlocalTWL}) is to
divide the computational domain into two parts:
 $I_1=[x_l,0]$ and $I_2=[0,x_r]$. Being given $\sigma$, in each interval an elliptic equation with Dirichlet boundary conditions is solved
 \begin{equation}\label{eq:uu1u2}
 \begin{array}{l}
 -\sigma \partial_xv_i=\partial_{xx}v_i+\mu v_i(1-\phi\star v), \qquad   v_i(0)=\epsilon, \qquad i=1,\; 2, \\ v_1(x_l)=1, \qquad v_2(x_r)=0.
 \end{array}
 \end{equation}
The convolution term is computed by defining $v$ as $v_1$ on $I_1$, $v_2$ and $I_2$, $1$ on $(-\infty,x_l)$ and $0$ on $(x_r,+\infty)$.

But the equation does not necessarily hold true at $x=0$. We define $\sigma^L$ so as to impose that the jump of derivatives at zero vanishes
\begin{equation}\label{eq:sigmafv}
\sigma^L=[\partial_xv_2(x_r)-\partial_xv_1(x_l)]+\int_{x_l}^{x_r}\mu v (1-\phi\star v)dx.
\end{equation}
\begin{lemma}\label{lemma:continuous}
When $(\sigma^L,v_1^L, v_2^L)$ satisfies (\ref{eq:uu1u2}) and (\ref{eq:sigmafv}) simultaneously, then $v^L$ is $C^1$ on $(x_l,x_r)$ and satisfies (\ref{eq:nonlocalTWL}).
\end{lemma}

We can write abstractly this problem as a  fixed point for a system of two equations
$(\sigma,v)= (\mathfrak{P}(v),\mathfrak{T}(\sigma))$.
It is straightforward to make it discrete using finite differences. The most efficient way to solve it is to use Newton iterations.

\section{The numerical results}
\label{sec:results}

\subsection{Convergence of the scheme}

The numerical results we present in this section are obtained with the hat function $\phi$ in (\ref{eq:phi2}), $\epsilon=0.1$ and we study the effect of the bifurcation parameter $\mu$. The diffusion term is treated implicitly by centered three point finite difference while the reaction term is put explicit.

In order to verify that the iterative scheme described in section \ref{sec:algorithm} converges to the right solution, a crucial quantity to look at is the truncation errors at zero
$$
E(0)=\frac{v^{L}_{1}-2v^{L}_{0}+v^{L}_{-1}}{\Delta x^2}+\sigma^{L}
\frac{v_1^{L}-v_{-1}^{L}}{2\Delta x}+v_0^{L}(1-(\phi\star v^L)_0).
$$ Here $v_{-1}^L,v_0^L,v_1^L$ are the values of $v^L$ at the grid points $-\Delta x,0,\Delta x$.
 The convergence results are displayed in Table \ref{tab:fisher2mu64}. One can see that $E(0)$ converges to zero as $\Delta x\to 0$, which shows that
our numerical results is a good approximation of (\ref{eq:nonlocalTWL}) on the whole computational domain. Note that better accuracy of $E(0)$ can be obtained if we use higher order numerical integration methods for the convolution term.
\begin{table}[h]
\begin{center}
\begin{tabular}{|c| c| c| c||c| c| c| c||c| c| c| c|}\hline
$L$&$\Delta x$&$|E(0)|$&$\sigma^L$&$L$&$\Delta x$&$|E(0)|$&$\sigma^L$&$L$&$\Delta x$&$|E(0)|$&$\sigma^L$
\\\hline
20&0.04&3.0619&15.3670&40&0.04&$3.0617$&15.3670&80&0.04&$3.0617$&15.3670\\\hline
20&$0.02$&1.7883&15.4930&40&0.02&$1.7905$&15.5028&80&0.02&$1.7888$&15.4930\\\hline
20&$0.01$&0.9721&15.5368&40&0.01&$0.9716$&15.5319&80&0.01&$0.9715$&15.5319\\\hline
20&$0.005$&0.5075&15.5429&40&0.005&$0.5075$&15.5429&80&0.005&$0.5075$&15.5429\\\hline
\end{tabular}\end{center}
\caption{\label{tab:fisher2mu64} Convergence of the truncation error at zero $E(0)$ and of the traveling velocity $\sigma^L$ for various values of $\Delta x$ and $L$, with kernel (\ref{eq:phi2}) and $\mu=64$. For $L=\infty$ the speed is $\sigma^\infty=2\sqrt{\mu}=16$.}
\end{table}


\subsection{Convergence of the traveling waves to $1$}

\begin{figure}
\begin{center}
\includegraphics[width=0.45\textwidth]{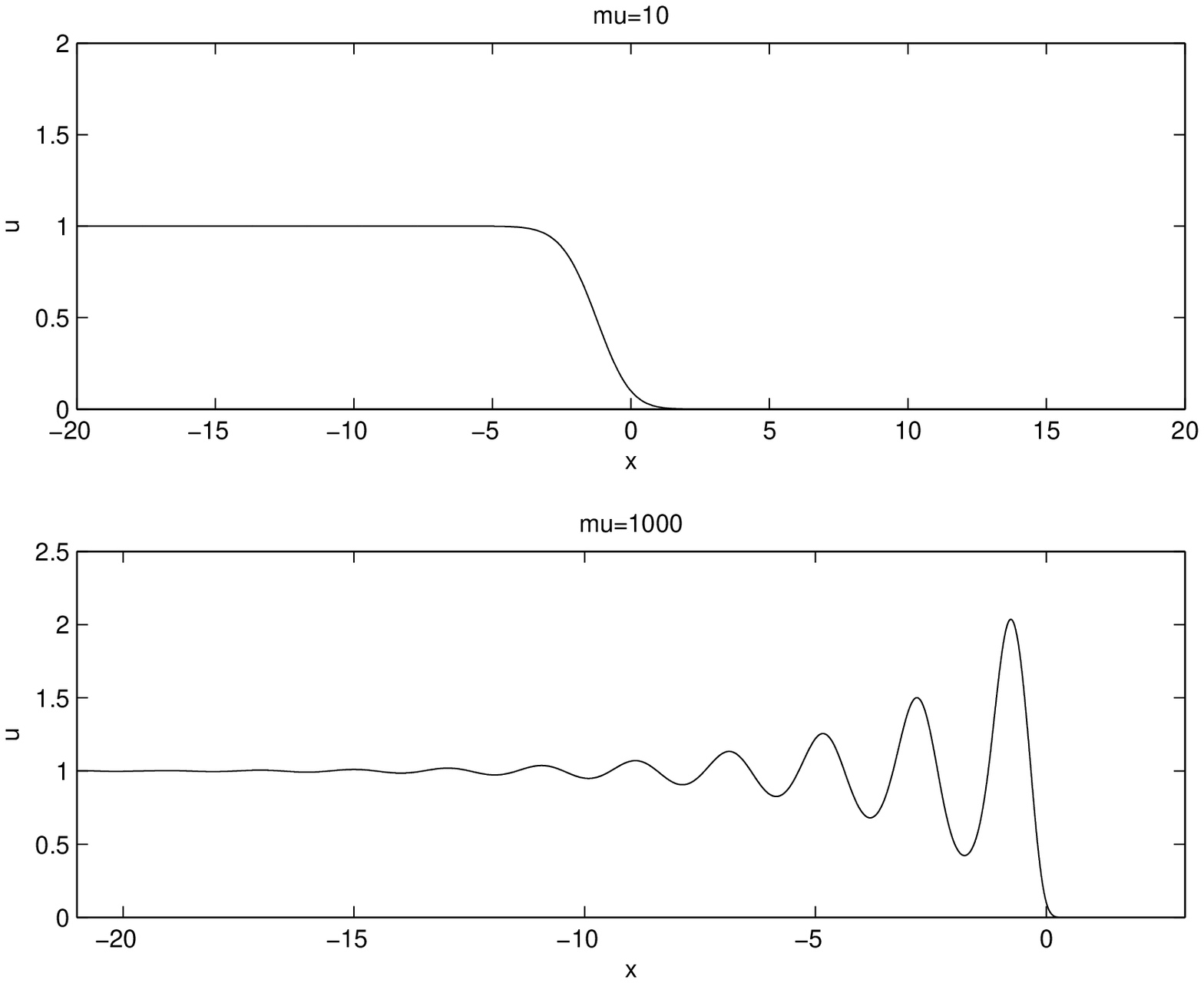}
 \includegraphics[width=0.45\textwidth]{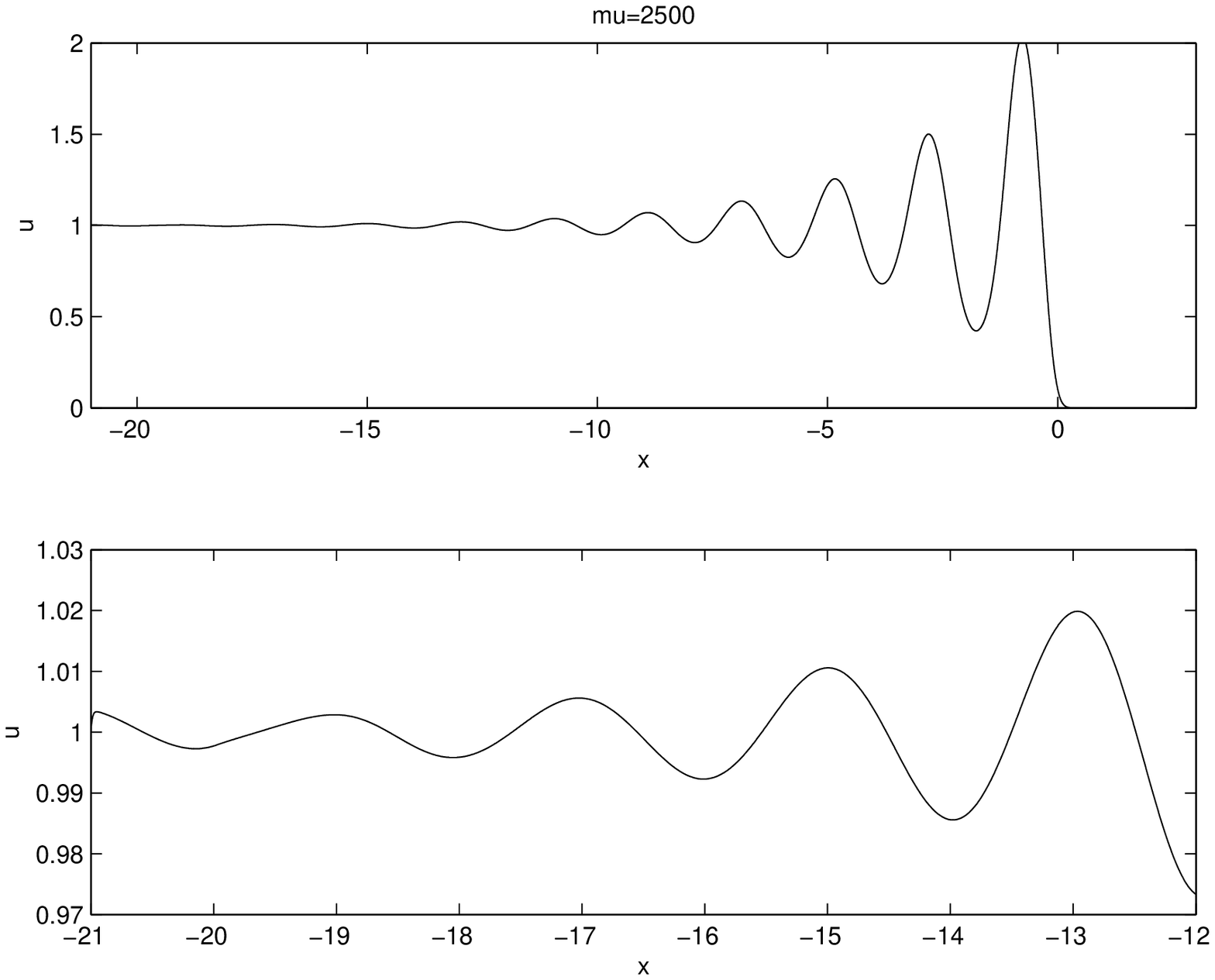}
\vspace{-4mm}
\caption{The traveling wave solution for the nonlocal Fisher equation (\ref{eq:nonlocalTW}) with kernel (\ref{eq:phi2}). Left: the numerical results for $\mu=1$ and $\mu=1000$. Right: the results for $\mu=2500$, the top subplot depicts $v$ while the bottom subplot is a zoom of the tail.}
\label{fig:nonlocal2}
\end{center}
\end{figure}

The traveling wave shapes for $\mu=10$, $\mu=1000$ and $ \mu=2500$ for kernel (\ref{eq:phi2}) are depicted in Figure \ref{fig:nonlocal2}.
When $\mu=10$, we observe a monotone traveling wave which connects $0$ to $1$, as for the local Fisher equation.
When $\mu$ grows, some oscillations appear.
Numerically, when we increase $L$, the amplitudes of the tail decrease and the bigger $\mu$ is, the slower the amplitudes decrease. The shapes of $v$ suggest that though $1$ is Turing unstable with the kernel (\ref{eq:phi2}) when $\mu=1000$ and $ \mu=2500$, the traveling waves will still connect $1$ to $0$.

We do not obtain the same type of structure than when we compute the solution of the evolution equation depicted in Figure \ref{fig:nonlocal1}. This means that there exist some traveling waves that connect $0$ to $1$, but that these waves do not attract the solution of the Cauchy problem associated with compactly supported initial data.

\subsection{Monotonicity of the traveling waves}

Lastly, we consider the critical value of $\mu$ for which the monotonicity of the traveling waves is broken.
Since the monotone traveling waves always connect $1$ to $0$, we perform the linearization, close to $x=-\infty$, by assuming $v\approx 1-e^{\lambda x}$ with $\lambda$ a real positive number
\footnote{We do not know if such a linearization is legitimate, but the technical arguments used in \cite{BNPR} to prove non-monotonicity for large $\mu$ are close to a linearization}.
After inserting this form into (\ref{eq:nonlocalTW}), using $\sigma=2\sqrt{\mu}$ and the smallness of $e^{\lambda x}$, $\lambda$ can be determined by $\mu$ through the equation
\begin{equation}
\frac{e^\lambda-e^{-\lambda}}{2\lambda}\mu-2\lambda\sqrt{\mu}-\lambda^2=0,
\label{eq:mulambda}
\end{equation}
a quadratic equation for $\sqrt{\mu}$, which gives
$\sqrt{\mu}=\frac{\lambda^2+\lambda\sqrt{\lambda^2+\lambda\sinh \lambda}}{\sinh \lambda}$. Thus
the critical $\mu$ that makes $\lambda$ no longer exist is $\mu_c=\sup_{\lambda>0}\big(\frac{\lambda^2+\lambda\sqrt{\lambda^2+\lambda\sinh \lambda}}{\sinh \lambda}\big)^2\approx 8.9$.

We have checked that this threshold $\mu_c\approx 8.9$ is correct on the numerical values. The maximum of
$v^L$ is $1$ when $\mu=9$, but exceeds slightly $1$ when $\mu=10$. In Figure \ref{fig:nonlocal2} with $\mu=10$, the wave is nearly monotonic, but, checking the numerical values, the maximum of $v$ is $1.0028$. This indicates that actually $v$ might be not monotone even for $L$ finite.

In \cite{BNPR} the authors have proved that, when $\mu>\mu_c$, the traveling wave is not monotone. One open question is to know if the traveling wave is monotone when $\mu<\mu_c$. Our simulation answers positively to this open question numerically.

\subsection{Traveling wave shapes for kernels in (\ref{eq:phinew})}
\begin{figure}
\begin{center}
\includegraphics[width=0.45\textwidth]{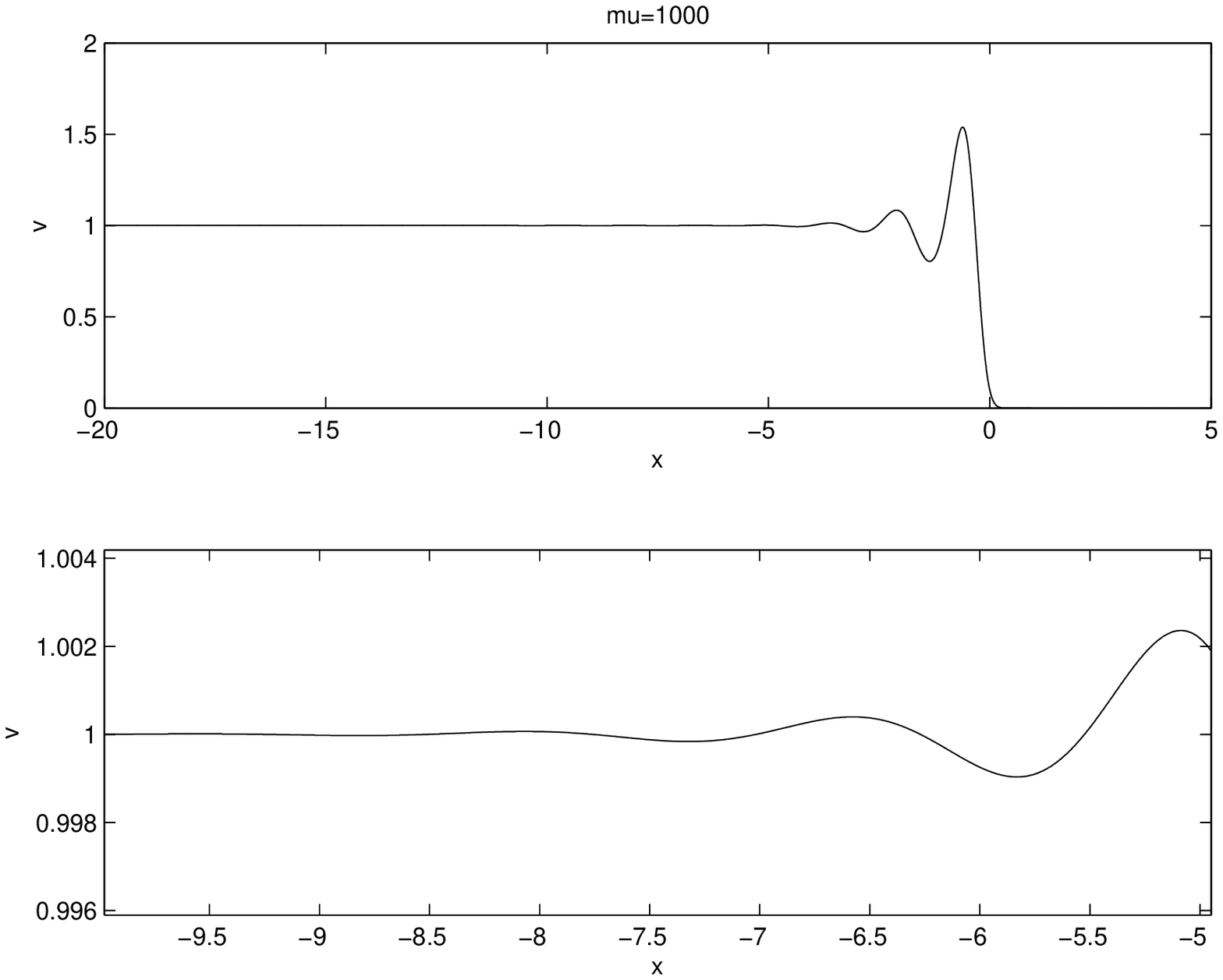}
\includegraphics[width=0.45\textwidth]{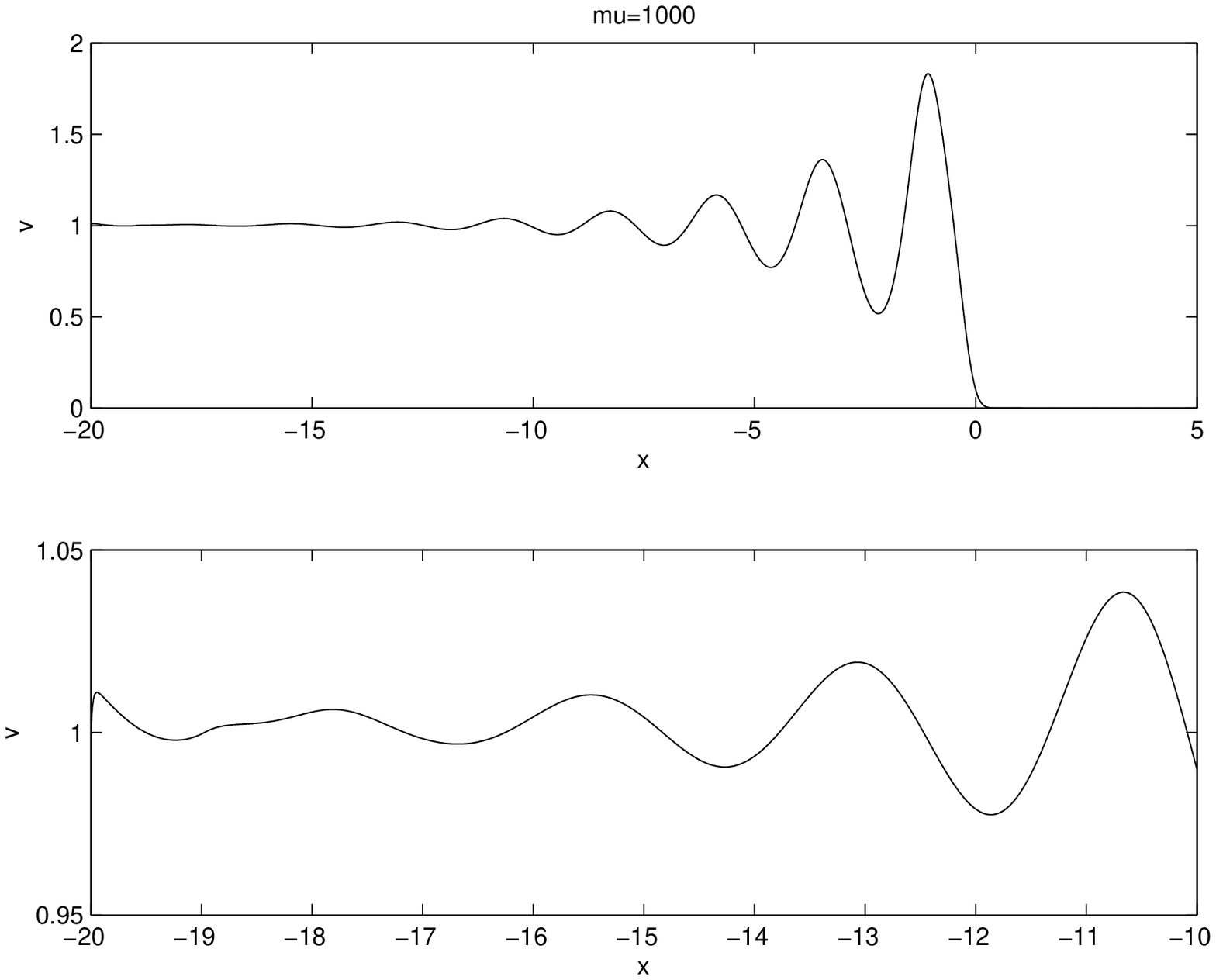}
\vspace{-4mm}
\caption{The traveling wave solution for the nonlocal Fisher equation (\ref{eq:nonlocalTW}) with $\mu=1000$ and kernels in (\ref{eq:phinew}). Left: the numerical results with kernel (\ref{eq:phinew1}). Right: the results for (\ref{eq:phinew2}). In all these pictures, the top subplot depicts $v$ while the bottom subplot is a zoom of the tail.}
\label{fig:newkernel}
\end{center}
\end{figure}

The numerical results with kernels in (\ref{eq:phinew}) are depicted in Figure \ref{fig:newkernel}.
With large $\mu=1000$, we can see similar phenomena such that some oscillations appear and
the waves become nonmonotone. For the two kernels in (\ref{eq:phinew}), the state $v\equiv1$ is Turing unstable as for (\ref{eq:phinew})
and Figure \ref{fig:newkernel} suggests again that the traveling waves will connect $1$ to $0$.

Finally, to complete the tests, we show the traveling wave for the Gaussian kernel
(\ref{eq:phi1}) for $\mu=1000$ in Figure \ref{fig:nonlocal1}.

\begin{figure}
\begin{center}
\includegraphics[width=0.45\textwidth]{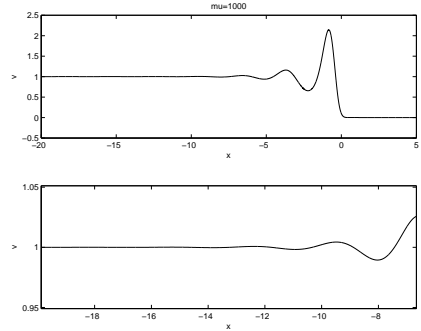}
\vspace{1mm}
\caption{The shapes of traveling wave solution for the nonlocal Fisher/KPP equation with Gaussian kernel (\ref{eq:phi1}) when $\mu=1000$.}\label{fig:nonlocal1}
\end{center}
\end{figure}


\begin{thebibliography}{00}

\bibitem{BH} H. Berestycki, F. Hamel, Front propagation in periodic excitable media, Comm. Pure Appl. Math. 55, 949--1032, 2002.

\bibitem{BNPR} H. Berestycki, G. Nadin, B. Perthame and L. Ryzhik, The non-local Fisher-Kpp equation: traveling waves and steady states. Nonlinearity (22), 2813--2844, 2009.

\bibitem{FKK} M.\ A. \ Fuentes, M. \ N.\ Kuperman and V.\ M. \ Kenkre, Nonlocal interaction effects on pattern formation in population dynamics,  Phys. Rev. Lett. 91(15),  15810414.1--15810414.4, 2003.

\bibitem{GVA} S. Genieys, V. Volpert and P. Auger, Pattern and waves for a model in population dynamics with nonlocal consumption of resources, Math. Modelling Nat. Phenom. 1, 65--82, 2006.

\bibitem{Gourley} S.\ A.\ Gourley, Travelling front solutions of a nonlocal Fisher equation,
J. Math. Biol. 41(3), 2000.

\bibitem{KPP} A.N. Kolmogorov, I.G. Petrovsky and N.S. Piskunov, Etude de l \'equation de la diffusion avec croissance de la
  quantit\'e de mati\`ere et son application \`a un probl\`eme biologique, Bulletin Universit\'e d'Etat \`a Moscou (Bjul. Moskowskogo Gos.
  Univ.), 1--26, 1937.

\bibitem{Shigesada} N. Shigesada, K. Kawasaki, E. Teramoto, Traveling periodic waves in heterogeneous environments, Theor. Population Biol. 30, 143--160, 1986.

		

\end{thebibliography}
\end{document}